# 利用金刚石氮-空位色心精确测量弱磁场的探索*

李路思,李红蕙,周黎黎,杨炙盛,艾清†

(北京师范大学物理学系, 北京 100875)

**摘 要**

本文基于金刚石氮-空位色心对精确测量微弱静磁场进行了探索.金刚石氮-空位色心电子自旋的退相干时间高度依赖于外磁场,而不同的退相干特征时间对磁场的灵敏度不同.本文通过对 NV 色心电子自旋在不同强度外磁场下的退相干过程进行模拟,得到不同退相干特征时间与磁场大小 $B$ 的高准确度关系,提出了基于响应度最高的退相干特征时间测量静态弱磁场大小和方向的方法,并分析了该方法测量静态弱磁场的灵敏度,证明该方法比一般测量磁场的仪器具有更高的灵敏度.

## 1 引 言

弱磁探测,如地磁场的精确测量,在航天、航空、航海导航以及远程精确制导等方面有重要意义.目前测量弱磁场的超灵敏度磁力仪有霍尔探针扫描显微镜、磁力显微镜[1,2]、质子磁力仪[2]、超导量子干涉仪[2,3]等,但需要在特定条件(如低温和高真空等)下进行,且成本较高.研究发现,鸟类的视网膜[4]、蝾螈的松果体[5]、蜜蜂的腹部[6]等器官内含有磁感应源."三叉神经假说"和"自由基对机制(RPM)"[7,8]是解释鸟类感磁原因的两个主流假说之一,文献[9]用"量子 Zeno 效应"解释

* 北京师范大学本科生科研训练与创新创业项目、国家自然科学基金青年基金(批准号:11505007)、清华大学低维量子物理国家重点实验室开放研究基金(批准号:KF201502)资助的课题.

† 通信作者. E-mail: aiqing@bnu.edu.cn 电话:010-58802817

了自由基对反应对于磁场的依赖性;文献[10]进一步探讨了自由基对的量子控制和纠缠的作用;受这一假说启发,文献[11]提出用推广的 Holstein 模型来描述这样一类依赖于自旋的化学反应,这对于弱磁场精确测量技术的发展具有重要借鉴意义.

金刚石-氮空位(NV)色心[12]的电子基态是一种自旋三重态系统,且具有易于初始化、易于读取、易于操控、相干时间长、常温操作等优点[13-16],这使得 NV 色心系统成为有望实现量子信息处理[17]和量子计算[18]的候选者之一. NV 色心系统的一个应用方向是作为高灵敏度探针进行弱磁场精确测量[19-21]. 已有研究[22]将 NV 色心系统用于蛋白质分子中的单个核自旋的弱磁测量,但对于静态弱磁场的精确测量则鲜少涉及. 文献[23]研究了 NV 色心退相干机制和周期性动态解耦控制对 NV 色心系统退相干行为的影响,同时表明,不同强度磁场下 NV 色心退相干的行为和时间不同,这启发了一种探测弱磁场的新方法.

本文将基于 NV 色心系统退相干时间[24,25]对外磁场大小和方向高度敏感的原理,探索一种能够精确测量微弱静磁场的新途径. 通过模拟 $^{13}C$ 原子含量为自然丰度、$^{14}N$ 高纯度的金刚石环境,实现 NV 色心电子自旋与周围 $^{13}C$ 核自旋的耦合,探究不同磁场大小下其耦合退相干的时间,确定退相干时间与外磁场大小之间的高灵敏度关系. 只要通过光学方法监测 NV 色心的退相干时间,就间接确定了 NV 主轴方向的磁场大小. 改变 NV 色心主轴方向,可以实现待测静态弱磁场的高精度三维测量.

## 2 NV 色心

### 2.1 NV 色心的物理性质

金刚石氮-空位色心,英文名为 Nitrogen-Vacancy (NV) Color Center,是金刚石

晶体中的一种点缺陷结构. 如图 1, 当金刚石中碳原子构成的面心立方晶格的一个碳原子被氮原子所取代, 且其近邻有一个晶格空位时, 就形成了 NV 色心. 氮原子与空位的连线方向即为 NV 主轴方向, 即[111]轴.

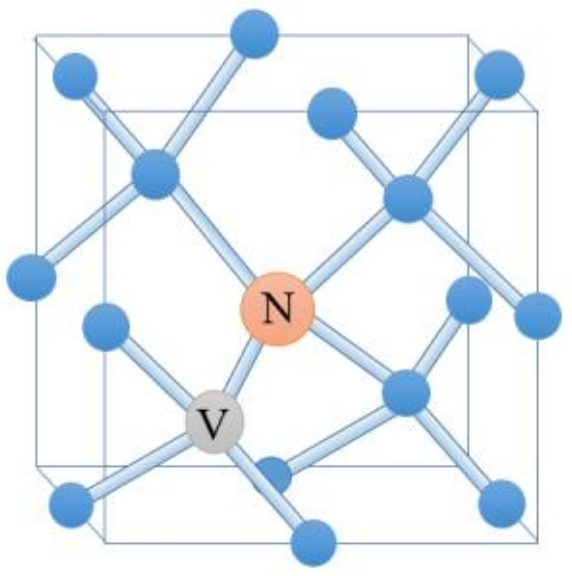

图 1　金刚石晶体中的氮-空位(NV)色心的物理结构. 其中红色为氮原子, 灰色表示空位.

Fig. 1. The physical structure of the NV center in diamond. The nitrogen atom and the vacancy are labeled by the red sphere and the gray sphere respectively.

基态的 NV 色心的总自旋为 S=1, 对应的能级结构和跃迁如图 2 所示. 实线表示可进行跃迁, 虚线表示不可进行跃迁, $\Gamma_s$和$\Gamma_d$表示无辐射跃迁. NV 的基态是自旋三重态$^3A$, $m_s=\pm1$和$m_s=0$之间存在$\Delta=2.87$GHz的零场劈裂. 由于存在声子相关的无辐射跃迁($\Gamma_s$和$\Gamma_d$), 故可用激光脉冲来极化电子自旋至$m_s=0$态, 即初始化. 电子由激发态跃迁到基态时发出荧光光子, 且由于跃迁过程不同, 初始态为$m_s=0$态的荧光强度比$m_s=\pm1$态的荧光强 20%~40%[26, 27], 由此可实现电子自旋状态的读出.

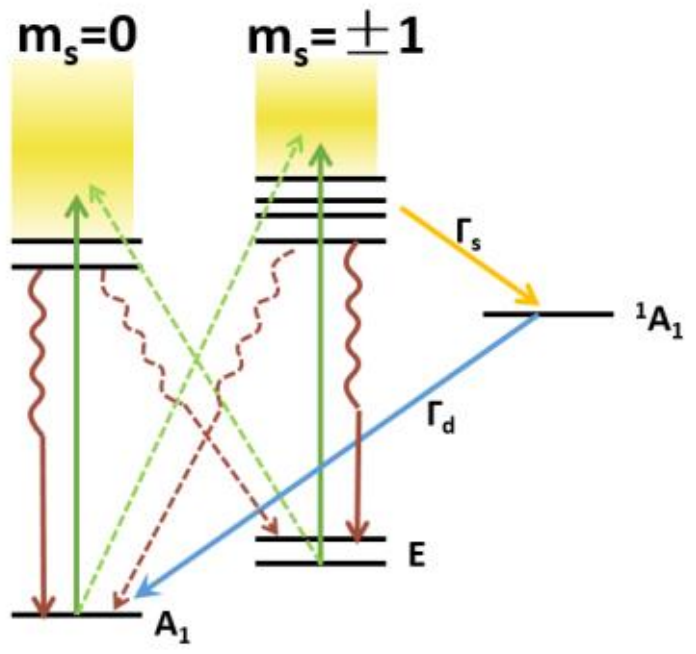

图 2　NV 色心的能级结构和跃迁.

Fig. 2. The electronic structure and transitions of the NV center.

2.2 NV 色心系统哈密顿量

金刚石中随机分布着自旋为 1/2 的 $^{13}$C, NV 色心电子自旋为 1, 会与 $^{13}$C 核自旋通过磁偶极相互作用耦合. 在外场 $\boldsymbol{B}$ 的作用下, 整个系统的哈密顿量为[28, 29]

$$H = H_{NV} + H_{bath} + H_{int} \tag{1}$$

其中 NV 的哈密顿量和 $^{13}$C 环境的哈密顿量为

$$H_{NV} = -\gamma_e \boldsymbol{B} \cdot \boldsymbol{S} + \Delta S_z^2 \tag{2}$$

$$H_{bath} = -\gamma_n \boldsymbol{B} \cdot \sum_i \boldsymbol{I}_i + H_{dip} \tag{3}$$

其中 $\gamma_e$、$\gamma_n$ 分别为电子自旋和核自旋的旋磁比. $H_{dip}$ 描述的是核自旋之间的磁偶极相互作用.

(1)式中最后一项

$$H_{int} = \boldsymbol{S} \cdot \sum_{\mathrm{i}} \boldsymbol{A}_i \boldsymbol{I}_{\mathrm{i}} \tag{4}$$

描述电子自旋与核自旋之间的相互作用, 是造成 NV 退相干的主要原因. 其中 $A_i$ 是第 i 个近邻核自旋与电子自旋之间的磁偶极相互作用.

故第 i 个核自旋的哈密顿量

$$H_i = -\gamma_n \boldsymbol{B} \cdot \boldsymbol{I}_{\mathrm{i}} + \boldsymbol{S} \cdot \boldsymbol{A}_i \boldsymbol{I}_{\mathrm{i}} \tag{5}$$

由于电子和核自旋之间大失谐, 超精细相互作用不会诱导电子自旋的翻转, 所以

当电子自旋处于$|m\rangle$态（$m=0,\pm1$为电子自旋的磁量子数)时，电子与核自旋的超精细耦合会对核自旋提供一个有效相互作用势, 即核自旋的有效哈密顿量为

$$H_i^{(m)} = -\gamma_n \boldsymbol{B}\cdot\boldsymbol{I}_{\mathbf{i}} + S_z^{(m)}\cdot\boldsymbol{A}_i\boldsymbol{I}_{\boldsymbol{i}} \tag{6}$$

此时第 i 个核自旋受到的有效磁场为

$$\boldsymbol{h}_i^{(m)} = \boldsymbol{B} - \boldsymbol{A}_i^{(m)}/\gamma_n \tag{7}$$

$$H_i^{(m)} = -\gamma_n \boldsymbol{h}_i^{(m)}\cdot\boldsymbol{I}_{\mathbf{i}} \tag{8}$$

其中$m=0,\pm1$为电子自旋的磁量子数, 也即

$$h_i^{(0)} = B \tag{9}$$

$$h_i^{(\pm1)} = B - A_i^{(\pm1)}/\gamma_n \tag{10}$$

核自旋绕该有效磁场进动, 并对电子自旋产生调制. 由于每个核所感受到的超精细耦合不同, 各个 $^{13}C$ 核自旋的进动频率也不同. 对系统施加 Hahn echo 脉冲, 可以有效消除这种差异, 使得各核自旋对于电子自旋的调制同时重相。但由于核自旋之间磁偶极相互作用的存在, 并不能通过对系统施加 Hahn echo 脉冲来消除影响,从而引起了 NV 电子自旋$|0\rangle$态和$|1\rangle$态之间的退相干[30].

# 3 NV 系统测量弱磁场

## 3.1 理论模拟

利用 MATLAB 对自然丰度(1. 1%) $^{13}C$ 环境中的 NV 色心系统在 Hahn echo 脉冲作用下的退相干过程进行模拟, 得到不同强度磁场大小下 NV 色心电子自旋$|0\rangle$态和$|1\rangle$态之间的退相干过程(如图 3). 在强、中、弱磁范围内, NV 色心退相干行为有显著差异. 该过程中有三个特征时间: $T_W$、$T_R$、$T_2$,其中, $T_W$为第一个半峰衰减到 1/e 的时间; $T_R$为相邻两峰之间的时间间隔, 是核自旋绕有效磁场进行拉莫进动的周期, 反映核自旋对于 NV 色心电子自旋相干性的调制[30]; $T_2$为各个峰的包络曲线衰减到 1/e 的时间, 反映系统退相干的快慢. NV 色心电子自旋的

相干性随时间演化可近似为

$$L(t)=\frac{1}{2}(1+\cos\frac{2\pi t}{T_R})e^{-t/T_2} \tag{11}$$

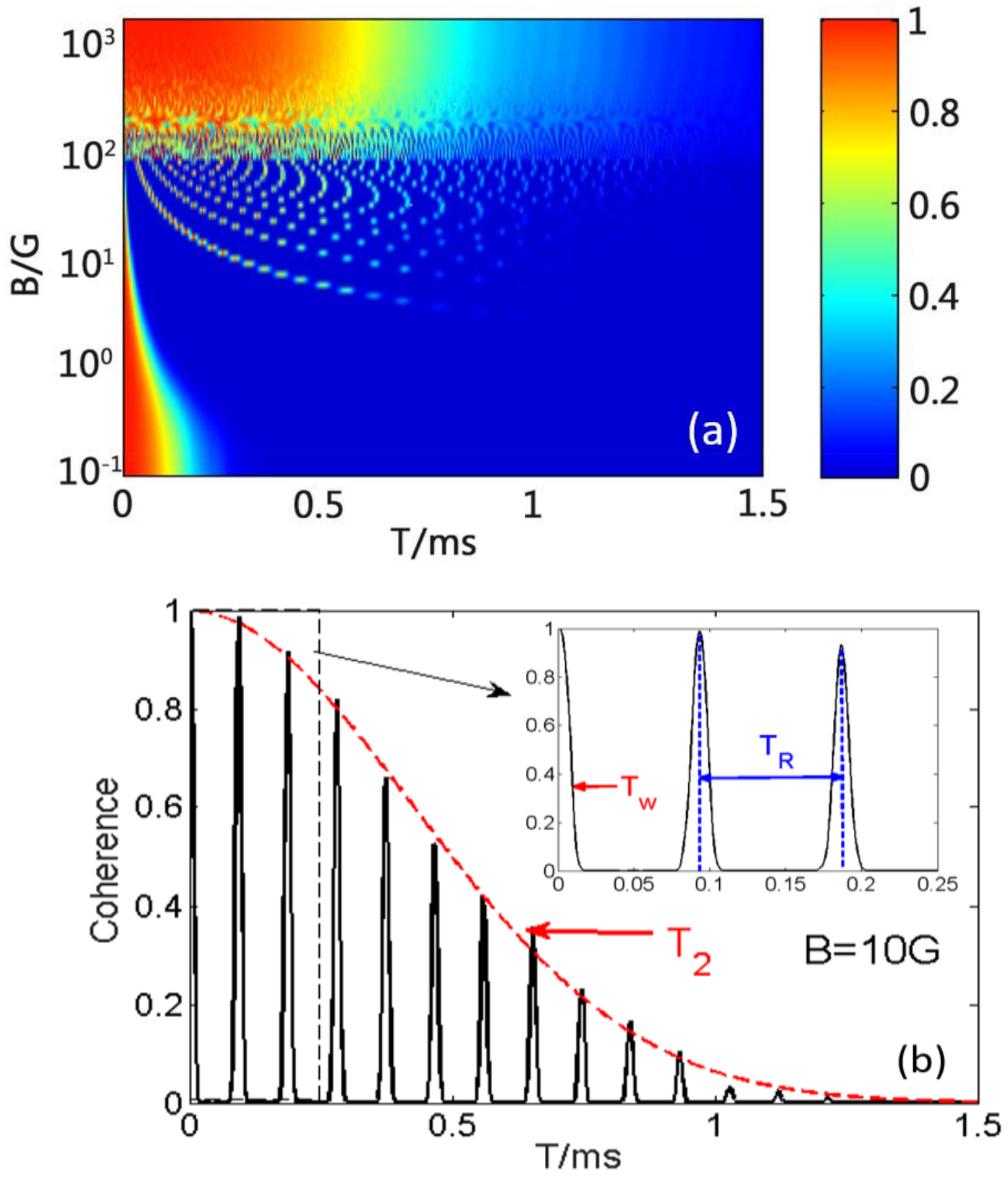

图 3 NV 色心电子自旋退相干过程模拟. (a) NV 色心$\left|0\right\rangle$态和$\left|1\right\rangle$态的相干性与相干时间 $T$ 和外磁场大小 $B$ 的关系; (b) $B$ =10 G 下 NV 色心的相干性随相干时间的变化图像, 小图为前三个峰的放大.

Fig. 3. Numerical simulation of the NV center electron spin decoherence: (a) Coherence between |0〉 and |+1〉 as a function of total evolution time and magnetic field; (b) Coherence as a function of evolution time under 10 G magnetic field. Inset: the first three peaks.

通过数据拟合, 得到$T_W$、$T_R$与外磁场 $B$ 的关系如下：

$$\begin{cases} T_R = 0.9366B^{-1} \\ T_W = 0.0427B^{-0.65} \end{cases} \tag{12}$$

由图 4 可知, 横向弛豫时间$T_2$随磁场增大整体上呈现增大趋势, 并在$B \geq 10\text{G}$时, 趋于一定值[16], 不适合用于指示磁场大小. $T_W$和$T_R$均随磁场增大而减小, 其中$T_W$在$1\sim100\text{G}$的磁场范围内与$B^{-0.65}$成正比, 在小磁场极限下趋向于定值; 在可测量范围内$T_R \propto 1/\gamma_n B$[23], 对磁场变化响应度较高. 若通过$T_R - B$的非线性关系

间接确定磁场大小, 可得到较高的精确度. 由于 $T_R \propto B^{-1}$, $T_R - B$ 曲线的斜率为 $\frac{\delta T_R}{\delta B} = -B^{-2}$, 即磁场越小, $T_R$ 对磁场大小的响应度越高.

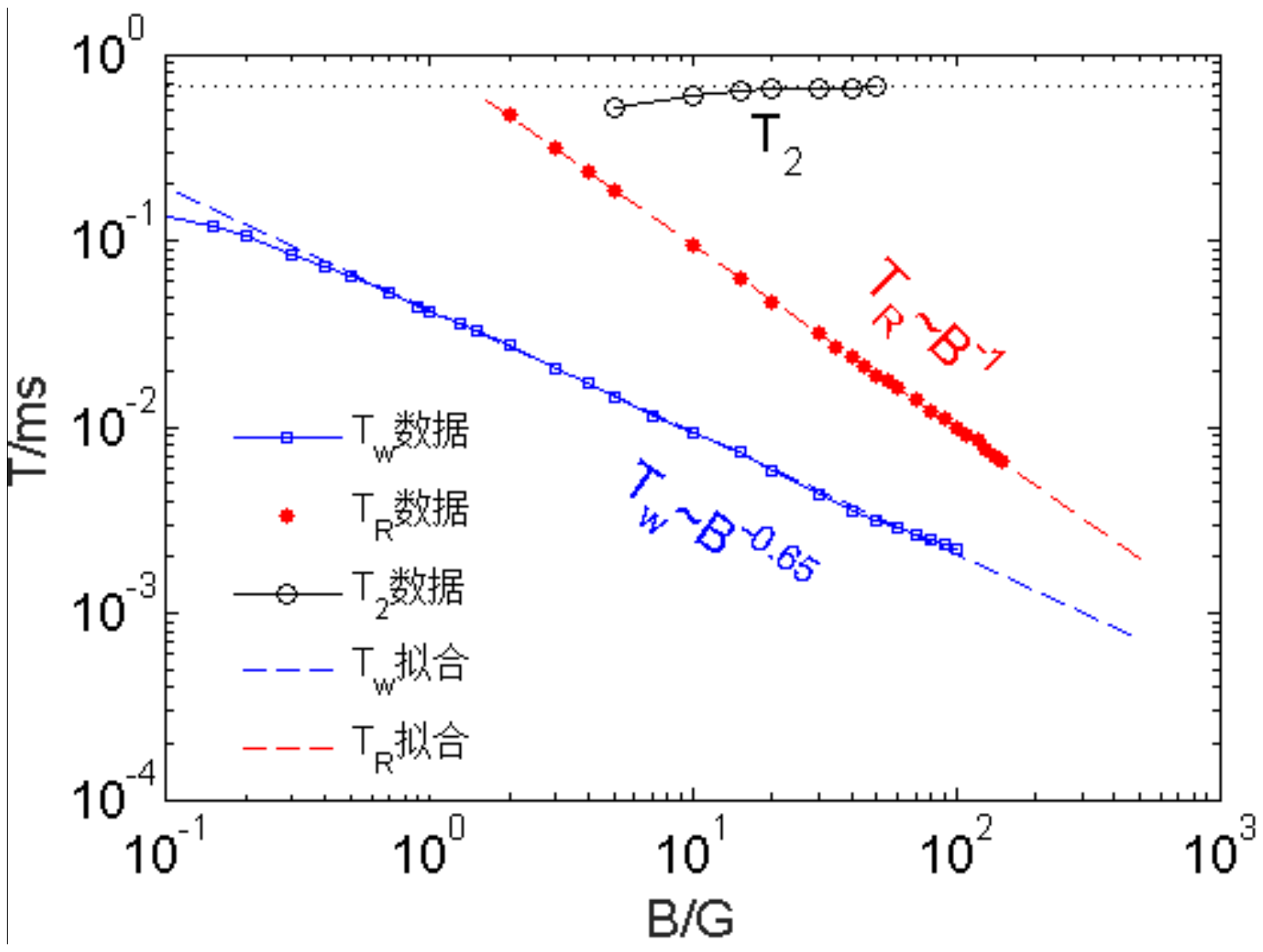

图 4 $^{13}C$ 自然丰度环境下, NV 色心电子自旋退相干的 $T_W$、$T_R$、$T_2$ 和外磁场大小的关系.

Fig. 4. The magnetic-field dependence of the three different timescales, $T_W$, $T_R$ and $T_2$ for NV center electron spin coherence in a $^{13}C$ bath with natural abundance.

图 4 也反映了利用 $T_R$ 测量静态弱磁场的磁场下限. 当外磁场较小时, $T_R$ 趋近于 $T_2$, 实际测量过程中可能出现第二个振荡出现之前系统就已经完全退相干的情况, 此时只有第一个半峰, 无法得到两峰间距 $T_R$. 此时可通过外加一个已知大小和方向的静磁场, 使总磁场处于可测量范围内.

## 3.2 确定磁场大小

已知 $T_R - B$ 的关系, 可以通过实验测得 $T_R$, 从而间接确定 NV 主轴方向的磁场大小. 为避免直接测量方向带来的不确定性, 可以使 NV 主轴分别处于三个相互正交的方向 $x, y, z$ 上, 测量 $T_R$ 并结合 $T_R - B$ 的非线性关系, 确定每次主轴方向的磁场大小 $B_x, B_y, B_z$, 则待测磁场 B 可表示为:

$$B=\sqrt{{B_x}^2+{B_y}^2+{B_z}^2} \tag{13}$$

磁场大小即可确定.

3.3 确定磁场方向

由于 $x,y,z$ 方向和各方向上磁场分量已知,磁场与 $x,y,z$ 方向余弦分别为

$$\cos\phi_i=\frac{B_i}{B}\ (i\text{为}x,y,z) \tag{14}$$

由此可以确定磁场方向. 由于对于每个磁场分量,方向有两个可能的取向,则磁场方向有六个可能取向. 为了确定磁场方向,可以先利用光学磁共振技术[31],观察 NV 色心在外磁场下发生塞曼分裂的情况.

存在外磁场 $B$ 时,忽略高阶小项,哈密顿量中与磁场 $B$ 有关的部分为

$$H_B=-\gamma_e\boldsymbol{B}\cdot\boldsymbol{S}+\Delta S_z^2-\gamma_n\boldsymbol{B}\cdot\sum_i\boldsymbol{I_i} \tag{15}$$

不考虑由 $^{13}C$ 和 $^{14}N$ 核自旋引起的超精细结构,则有

$$H_B^{'}=-\gamma_e\boldsymbol{B}\cdot\boldsymbol{S}+\Delta S_z^2 \tag{16}$$

在 $S_z$ 表象下, $\hat{S}_x$、$\hat{S}_y$、$\hat{S}_z$ 的矩阵元分别如下(因子 $\hbar$ 略去未记)

$$S_x=\begin{pmatrix}0&\frac{1}{\sqrt{2}}&0\\ \frac{1}{\sqrt{2}}&0&\frac{1}{\sqrt{2}}\\ 0&\frac{1}{\sqrt{2}}&0\end{pmatrix},\ S_y=\begin{pmatrix}0&\frac{-i}{\sqrt{2}}&0\\ \frac{i}{\sqrt{2}}&0&\frac{-i}{\sqrt{2}}\\ 0&\frac{i}{\sqrt{2}}&0\end{pmatrix},\ S_z=\begin{pmatrix}1&0&0\\0&0&0\\0&0&-1\end{pmatrix} \tag{17}$$

由定态薛定谔方程解能量本征值 $E$ 得方程

$$x^3+2\Delta' x^2+(\Delta'^2-2B^2)x-2(B^2-B_z^2)\Delta'=0 \tag{18}$$

其中

$$\begin{cases} x = \frac{\sqrt{2}}{\gamma_e} E \\ \Delta' = \frac{\Delta}{2\sqrt{2}\gamma_e} \\ B = \sqrt{B_x{}^2 + B_y{}^2 + B_z{}^2} \end{cases} \tag{19}$$

若外界磁场 $B$ 方向沿 NV 主轴方向, 即 $B = B_z$ , 则(17)式方程的解为:

$$\begin{cases} E_0 = 0 \\ E_1 = \Delta + \gamma_e B \\ E_{-1} = \Delta - \gamma_e B \end{cases} \tag{20}$$

此时从 $|0\rangle$ 态向 $|1\rangle$ 态与 $|\text{-}1\rangle$ 态跃迁的所吸收或放出的光子频率分别为 $\Delta \pm \gamma_e B$ . 当 NV 主轴与外磁场方向相同时, 对该 NV 色心扫光磁共振谱, 理论上有如下结果:

a. 两个共振峰关于 $\Delta = 2.87\text{GHz}$ 对称;

b. 两共振峰之间的频率差为 $\delta = 2\gamma_e B$ .

由此可以确定外磁场方向.

3.4 测量灵敏度

考虑单个 NV 色心系统, 电子自旋初态为 $|0\rangle$ 态, 对其施加 $\frac{\pi}{2} - \frac{\tau}{2} - \pi - \frac{\tau}{2} - \frac{\pi}{2}$ 的微波脉冲序列, 经光探测得到荧光信号

$$S \propto P_0(\tau) = \frac{1}{4}[2 + L(\tau) + L^*(\tau)]$$

$$= \frac{1}{2} + \frac{1}{4}(1 + \cos\frac{2\pi\tau}{T_R})e^{-\tau/T_2} \tag{21}$$

其中 $P_0(\tau)$ 是系统末态中 $|0\rangle$ 的布居数。

由图 4 知, $T_R$ 、 $T_2$ 的值均与外磁场 $B$ 有关. 其中, $T_2$ 对外磁场 $B$ 变化的响应度远小于 $T_R$ ，因此，外磁场 $B$ 变化一个小量 $\delta B$ 时，可将 $T_2$ 视为一常数，则由外磁场变化引起的荧光信号变化为

$$\delta S = \frac{\partial S}{\partial B}\delta B$$

$$=\frac{1}{4}e^{-\tau/T_2}\sin(\frac{2\pi\tau}{T_R})(-\frac{2\pi\tau}{T_R^2})\frac{\partial T_R}{\partial B}\delta B$$

$$=\frac{1}{4}e^{-\tau/T_2}\sin(\frac{2\pi\tau}{T_R})\frac{2\pi\tau}{\frac{\alpha^2}{B^2}}\frac{\alpha}{B^2}\delta B$$

$$=\frac{\pi\tau}{2\alpha}e^{-\tau/T_2}\sin(\frac{2\pi\tau}{T_R})\delta B \qquad (22)$$

其中运用了（12）式 $T_R=\alpha B^{-1}$，$\alpha=0.9366\mathrm{ms}\cdot\mathrm{G}$.

同时, 由于荧光读出时光子散射噪声引起的读出噪声

$$\delta S = N^{-1/2}/C \qquad (23)$$

其中 $C$ 为一与收集效率有关的常数，收集效率为 5%时，$C\approx 0.3$[32]; $N=\frac{T}{\tau}$ 为测量次数, $T$ 为总的测量间隔， $\tau$ 为积分时间. 可测量的最小磁场改变量

$$\delta B=\frac{1}{\frac{\pi C}{2\alpha}e^{-\tau/T_2}\sin(\frac{2\pi\tau}{T_R})\sqrt{T\tau}} \qquad (24)$$

由此定义灵敏度[32]

$$\eta\equiv\delta B\sqrt{T}=\frac{2\alpha}{\pi C|\sin(\frac{2\pi\tau}{T_R})|\sqrt{\tau}}e^{\tau/T_2} \qquad (25)$$

当满足 $\frac{2\pi\tau}{T_R}=(k+\frac{1}{2})\pi(k=0,1,\cdots)$ 且 $\tau\approx T_2/2$ 时，灵敏度达到最小值

$$\eta\approx\frac{2\alpha}{\pi C}\sqrt{\frac{2e}{T_2}} \qquad (26)$$

在本文采用的 $^{13}$C 自然丰度的单个 NV 系统中,退相干时间 $T_2\approx 0.5\mathrm{ms}$,此时 $\eta\approx 20\mu\mathrm{T/Hz^{1/2}}$.对于含 n 个 NV 色心的集群样品,由于荧光信号被放大 n 倍,增大了信噪比,可使得灵敏度提高 $\sqrt{n}$ 倍. 同时,由(22)式可知延长退相干时间也可以提高灵敏度.在单个 NV 色心系统中,电子自旋与 $^{13}$C 核自旋之间的相互作用是造成 NV 退相干的主要原因. 如图 5 所示，随着 $^{13}$C 浓度增大, 核与核之间的磁偶极相互作用造成各核自旋拉莫进动的不同步，进而导致电子自旋退相干更加剧烈,

退相干时间 T2 缩短. 因此采用合适的脉冲序列和经 $^{12}C$ 纯化的金刚石样品延长退相干时间[33]，从而显著提高测量弱磁场的灵敏度. 目前对于自然 $^{13}C$ 丰度的样品 $T_2$ 已可达 0.6s[34],另外可以通过同位素纯化，将 $^{13}C$ 浓度降低到万分之一以内[33]此时 $\eta \approx 60\text{nT/Hz}^{1/2}$.

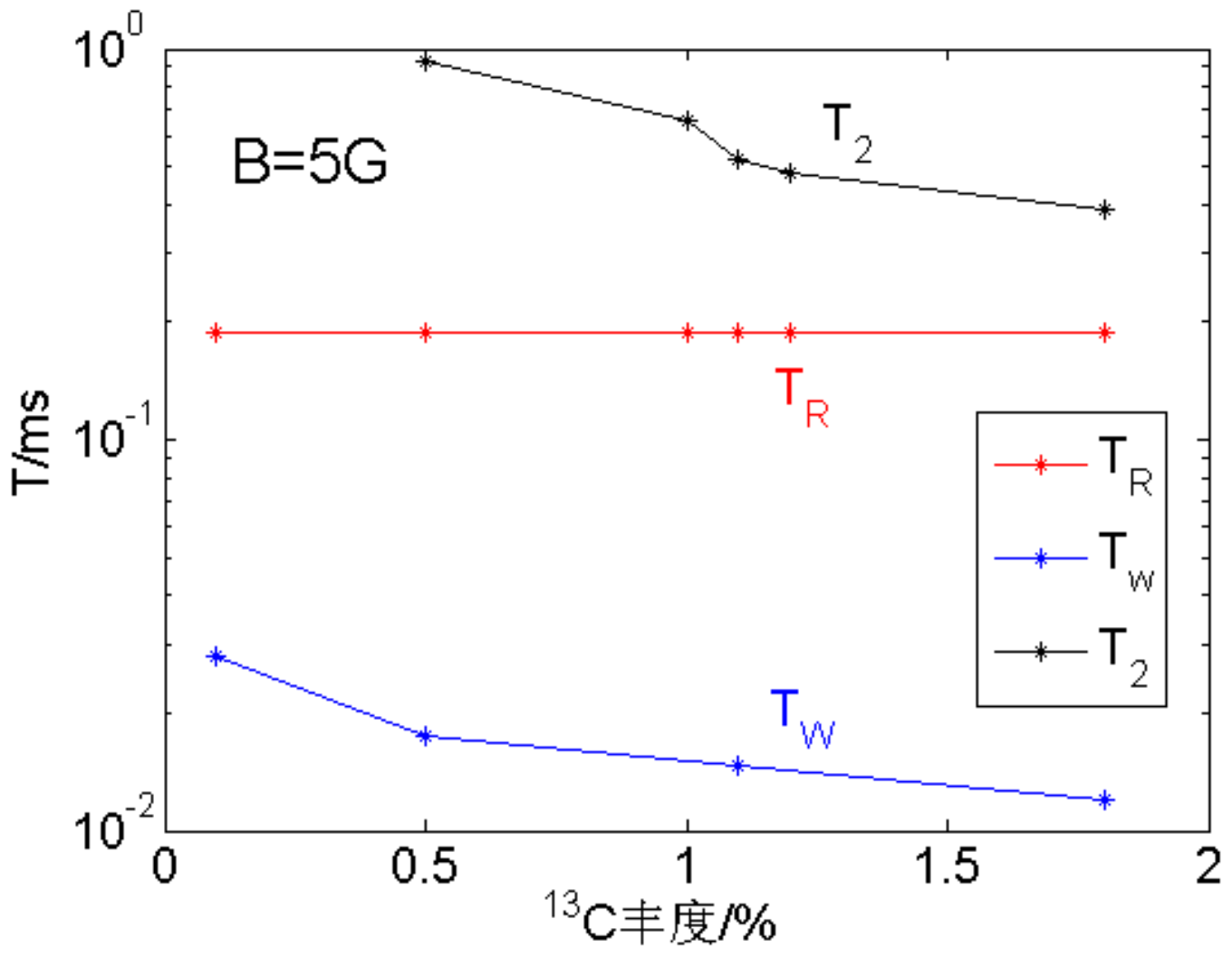

图 5 $^{13}C$ 丰度对 $T_W$、$T_R$、$T_2$ 的影响.

Fig. 5. The effect of $^{13}C$ abundance on the timescales $T_W$、$T_R$、$T_2$.

表 1 几种主要的磁力计及其灵敏度

Table 1. Sensitivities of different magnetometers.

| 名称 | 灵敏度 |
|---|---|
| 霍尔探针扫描显微镜[1,2] | $80\mu\text{T/ Hz}^{1/2}$ |
| 磁力显微镜[1,2] | $50\text{nT/Hz}^{1/2}$ |
| 质子磁力仪[2] | $0.1\text{nT/Hz}^{1/2}$ |
| 超导量子干涉仪[2,3] | $1\text{fT/ Hz}^{1/2}$ |

为了对比 NV 色心和其它磁力计测量微弱磁场的精度, 我们在表 1 中列举了几种主要磁力计的灵敏度. 超导量子干涉仪的灵敏度虽然高于本文测弱磁方法, 但同时需要低温(4K)工作条件,对于室外勘测不利. 本文方法灵敏度已足以进行地磁场量级的弱磁场测量，且可通过施加脉冲优化和同位素纯化的方法进一步提

高测量灵敏度.

## 4 结 论

地磁约为0.5G，属于弱磁范围内，在人类、动物生活、科研事业研究[36]等方面有巨大应用. 本文借鉴鸟类感磁机制，提出了一种利用量子相干系统精确测量静态弱磁场的方法. 通过哈密顿量求解，简要分析了金刚石中 NV 色心电子自旋退相干的原因，然后模拟了高纯度金刚石中单个 NV 色心与周围 $^{13}C$ 核自旋热库的耦合，得到了电子自旋 $|0\rangle$ 态与 $|1\rangle$ 态的相干性随演化时间和外部磁场大小的分布图，进而得到退相干特征时间 $T_W$、$T_R$、$T_2$ 与外部磁场 $B$ 的幂函数形式的高精确度关系. 通过分析 $T_W$、$T_R$、$T_2$ 分别对外磁场 $B$ 的响应度，确定将 $T_R$ 用于本方法中指示外磁场的“标尺”，即对于一个特定的 NV 色心系统，电子自旋退相干过程的 $T_R$ 与外磁场 $B$ 有严格的对应关系. 只要在光学磁共振实验中确定 $T_R$，由 $T_R$-$B$ 曲线即可得到该环境下 NV 主轴方向的静磁场分量，通过在相互正交方向上的三次测量，即可确定外磁场的三维分布. 对于该方法造成的磁场方向的不确定性，可利用连续的光磁共振谱实验解决. NV 色心主轴方向与外磁场方向平行时，塞曼分裂的连续光磁共振谱将关于 $\Delta = 2.87\mathrm{GHz}$ 对称，通过调整 NV 色心指向，可大致确定磁场方向，由此消除该方法测量磁场方向的不确定性. 通过灵敏度分析，本方法对于静态弱磁场的测量灵敏度可达 $60\mathrm{nT/Hz}^{1/2}$，且可通过增大样品中 NV 色心浓度和同位素纯化的方法达到更高的灵敏度.

参考文献:

# Measurement of weak static magnetic fields with nitrogen-vacancy color center*

Li Lu-Si Li Hong-Hui Zhou Li-Li Yang Zhi-Sheng Ai Qing†

(*Department of Physics, Beijing Normal University, Beijing 100875, China*)

Abstract

We propose a strategy to measure weak static magnetic fields with nitrogen-vacancy color center in diamond. Inspired by avian magnetoreception models, we consider the feasibility of utilizing quantum coherence phenomena to measure weak static magnetic fields. Nitrogen-vacancy (NV) color centers are regarded as the ideal platform to study quantum sciences as a result of its long coherence time up to a millisecond timescale. In high-purity diamond, hyperfine interaction with $^{13}C$ nuclear spins dominates the decoherence process. In this paper, we numerically simulate the decoherence process between $|0\rangle$ and $|+1\rangle$ of the individual NV color center spin in $^{13}C$ nuclear baths with various of magnitudes of external magnetic fields. By applying Hahn echo into the system, we obtain the coherence of NV color center spin as a function of total evolution time and magnetic field. Furthermore we obtain the high-accuracy relationship

* Project supported by the Undergraduate Research Foundation of Beijing Normal University, and the National Natural Science Foundation of China under Grant No. 11505007, and the Open Research Fund Program of the State Key Laboratory of Low-Dimensional Quantum Physics, Tsinghua University Grant No. KF201502..

† Corresponding author. E-mail: aiqing@bnu.edu.cn Telephone：010-58802817

between the three decoherence-characteristic timescales, i.e. $T_W$、$T_R$、$T_2$, and magnetic field $B$. And we draw a conclusion that $T_R$ has the highest sensitivity about magnetic field among the three time-scales. Thus, for a certain NV color center, $T_R$ can be the scale for the magnitude of magnetic field, or rather, the component along the NV electronic spin axis. When measuring an unknown magnetic field, we adjust the NV axis to three mutually orthogonal directions respectively. By this means, we obtain the three components of the magnetic field and thus the magnitude and direction of the actual magnetic field. The accuracy could reach 60nT/Hz$^{1/2}$,and could be greatly improved by using an ensemble of NV color centers or diamond crystals purified with $^{12}$C atoms.